\documentclass[
aps,
prx,
superscriptaddress,
twocolumn,
a4paper,
10pt,
floatfix,
longbibliography
]{revtex4-2}

\usepackage[american]{babel}

\usepackage{grffile}

\usepackage{newtxtext,newtxmath}
\usepackage{microtype}

\usepackage{amsmath}
\usepackage{amssymb}
\usepackage{bbm}

\usepackage{mathtools}
\usepackage{dsfont}
\usepackage{braket}
\usepackage{cancel}
\usepackage{slashed}

\usepackage{graphicx}
\usepackage{subcaption}
\usepackage[svgnames]{xcolor}

\usepackage{siunitx}

\usepackage{ragged2e}
\usepackage{array}
\usepackage{tabularx}
\usepackage{booktabs}

\definecolor{cset-aps-blueberry}{RGB}{28,128,158}
\definecolor{cset-aps-blue}{RGB}{46,44,184}
\definecolor{cset-aps-turquoise}{RGB}{0,67,88}
\definecolor{cset-aps-limegreen}{RGB}{190,219,67}
\definecolor{cset-aps-green}{RGB}{31,138,112}
\definecolor{cset-aps-yellow}{RGB}{255,225,25}
\definecolor{cset-aps-orange}{RGB}{253,116,0}
\definecolor{cset-aps-red}{RGB}{219,0,43}

\usepackage{tikz}
\usetikzlibrary{external}
\usepackage{pgfplots}
\pgfplotsset{%
	every axis legend/.append style={%
		cells={anchor=west},
		at={(0.96,0.04)},
		anchor=south east,
		font=\scriptsize,
	},
	every axis/.append style={%
		yticklabel style={%
			/pgf/number format/fixed zerofill,
			/pgf/number format/precision=2},
	},
	width= \textwidth,
	height=8cm,
	xmajorgrids=true,
	xminorgrids=false,
	minor x tick num=1,
}

\usepackage{pict2e,picture}

\makeatletter
\DeclareRobustCommand{\Arrow}[1][]{%
	\check@mathfonts
	\if\relax\detokenize{#1}\relax
	\settowidth{\dimen@}{$\m@th\rightarrow$}%
	\else
	\setlength{\dimen@}{#1}%
	\fi
	\sbox\z@{\usefont{U}{lasy}{m}{n}\symbol{41}}%
	\begin{picture}(\dimen@,\ht\z@)
		\roundcap
		\put(\dimexpr\dimen@-.7\wd\z@,0){\usebox\z@}
		\put(0,\fontdimen22\textfont2){\line(1,0){\dimen@}}
	\end{picture}%
}
\makeatother

\usepackage{hyperref}
\hypersetup{%
	colorlinks=true,
	linkcolor={cset-aps-red},
	linkbordercolor={cset-aps-red},
	filecolor={cset-aps-orange},
	filebordercolor={cset-aps-orange},
	citecolor={cset-aps-blue},
	citebordercolor={cset-aps-blue},
	urlcolor={cset-aps-green},
	urlbordercolor={cset-aps-green},
	menucolor={cset-aps-limegreen},
	menubordercolor={cset-aps-limegreen},
	breaklinks=true,
	pdfborderstyle={/S/U/W 2},
	pdfpagemode=UseOutlines,
	pdfstartpage={1},
}

\newcommand{\ee}{\text{e}}
\newcommand{\ii}{\text{i}}
\newcommand{\dd}{\mathrm{d}}
\newcommand{\sinc}{\mathrm{sinc}}

\usepackage{bm}
\renewcommand{\vec}[1]{\bm{#1}}

\newcommand{\affULM}{\address{Institut f{\"u}r Quantenphysik and Center for Integrated Quantum Science and Technology (IQST), Universit{\"a}t Ulm, Albert-Einstein-Allee 11, D-89081 Ulm, Germany}}
\newcommand{\affDLR}{\address{German Aerospace Center (DLR), Institute of Quantum Technologies, 89081 Ulm, Germany}}

\newcommand{\orcid}[1]{\href{https://orcid.org/#1}{\includegraphics[width=7pt]{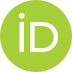}}}

\usepackage{comment}

\usepackage{lipsum}

\usepackage[normalem]{ulem}

\pgfplotsset{compat=1.16}

\makeatletter
\long\def\@makecaption#1#2{%
	\vskip\abovecaptionskip
	\setbox\@tempboxa\hbox{\small #1: #2}%
	\ifdim \wd\@tempboxa >\hsize
	\parbox[t]{\hsize}{\small \justifying #1: #2}%
	\else
	\parbox[t]{\hsize}{\small \justifying #1: #2}%
	\fi
	\vskip\belowcaptionskip
}
\makeatother

\begin{document}
	
	\title{Signatures of the circular Unruh effect in electric and magnetic dipole transitions of multilevel atoms} 
	\collaboration{This article has been published in \href{https://doi.org/10.1103/6pz4-wncx}{Phys. Rev. Res. \textbf{8}, 043029 (2026)} under the terms of the \href{https://creativecommons.org/licenses/by/4.0/}{Creative Commons Attribution License 4.0 [CC BY]}}
	
	\author{Gregor Janson~\orcid{0009-0002-7509-4550}}
	\email{gregor.janson@uni-ulm.de}
	\affULM
	
	\author{Fabio Di Pumpo~\orcid{0000-0002-6304-6183}}
	\email{fabio.di-pumpo@uni-ulm.de, fabio.di-pumpo@gmx.de}
	\affULM
	
	\author{Lorenz Thoma~\orcid{0009-0000-7509-3748}}
	\affULM
	
	\author{Maxim A. Efremov~\orcid{0000-0001-6395-9663}}
	\affULM
	\affDLR
	
	\begin{abstract}
		The circular Unruh effect is the excitation of a detector moving along a planar circular trajectory within an electromagnetic vacuum.
		We demonstrate that the magnetic dipole transitions in an atom, acting as the detector, dominate the electric dipole transitions.
		Our analysis of both free-space and cavity schemes shows that the sensitivity to the circular Unruh effect can be maximized by balancing the minimization of mode volume against the resulting decrease in mode density.
		Moreover, we propose a novel measurement scheme that uses the atom's multilevel structure to suppress the spontaneous emission rate, thereby enabling the experimental detection of the circular Unruh effect.
	\end{abstract}
	
	\maketitle
	
	\section{Introduction}\label{sec: Introduction}
	\label{sec.Introduction}
	Quantum field theory is usually formulated in inertial flat-spacetime Minkowski reference frames~\cite{PeskinSchroeder1995}.
	In this setting, the existence of a global, timelike Killing vector field is necessary to define a unique vacuum state of a quantum field, thereby providing a global definition of excitations that remains invariant when transforming between inertial frames.
	Consequently, each inertial flat-spacetime observer detects the same Minkowski vacuum, encoding the zero-excitation state~\cite{Misner1973, Wald1984}.
	
	However, for non-inertial reference frames~\cite{Misner1973, Wald1984}, arising {\it e.g.} from gravitational fields~\cite{Parker1969, Hawking1975, BirrellDavies1982} or non-inertial motion~\cite{Fulling1973, Davies1975, Unruh1976, Scully2019, Sudhir2021, Letaw1981, Crispino2008, Bell1983, Hacyan1986, Bell1987, Kim1987, Levin1993, Audretsch1995, Davies1996, Unruh1998, Rosu2005, Lochan2020, Arya2023, Zhou2025, Zheng2025, Parry2025}, the number of excitations is generally not invariant when transforming from one frame to another.
	A prominent example of this phenomenon is the linear Unruh effect~\cite{Fulling1973, Davies1975, Unruh1976, Letaw1981, Crispino2008, Scully2019, Sudhir2021}, where the Minkowski vacuum appears as a thermal excited state to an observer undergoing constant proper acceleration $a$.
	As such, it has been shown that an accelerated detector, {\it e.g.} built from massive particles or atoms~\cite{Unruh1976, Letaw1981, Crispino2008, Scully2019, Sudhir2021}, interacting with a quantized electromagnetic field measures a non-vanishing excitation rate $\Gamma_\text{lin} \propto \left\{\exp[\Delta E/(k_{\rm B} T_{\rm U})]-1\right\}^{-1}$, even if this field is in its vacuum state in an inertial frame. 
	Here, $\Delta E$ is the energy spacing of the two-level detector and $T_{\rm U} = \hbar a/(2\pi c k_{\rm B})$ denotes the Unruh temperature, and $k_{\rm B}$ the Boltzmann constant. 
	
	In this work, we consider a multilevel atom measuring the Unruh effect in circular setups~\cite{Letaw1980, Letaw1981, Crispino2008, Bell1983, Hacyan1986, Bell1987, Kim1987, Levin1993, Audretsch1995, Davies1996, Unruh1998, Rosu2005, Lochan2020, Arya2023, Zhou2025, Zheng2025, Parry2025}, i.\,e., the circular Unruh effect. 
	Here, instead of a linear acceleration, we consider a rotating multilevel atom following a classical center-of-mass trajectory, where the acceleration vector continuously rotates while pointing toward the center of the circular path.
	This atom acts as a detector for a quantized electromagnetic background field in a vacuum state, which is defined in the (inertial) laboratory frame.
	While in this frame an observer would measure a vanishing number of excitations, in the reference frame co-rotating with the atom, the field vacuum appears as an excited state, resulting in a non-vanishing atomic excitation rate.
	
	In contrast to the linear Unruh effect, the excitation rate of a circularly moving detector does not strictly assume the form of a thermal spectrum~\cite{Letaw1980, Bell1983, Bell1987}. 
	More precisely, an effective temperature \mbox{$T_\text{eff}(\Delta E) = \Delta E/\{k_B \ln[\Gamma_-(\Delta E)/\Gamma_+(\Delta E)]\}$} is often introduced via Einstein's detailed-balance condition comparing the excitation rate $\Gamma_+$ and the spontaneous emission rate $\Gamma_-$~\cite{Crispino2008, Biermann2020,Zhou2025}.
	It has been shown~\cite{Biermann2020,Zhou2025} that the effective temperature for circular motion, unlike for linear acceleration, depends on the energy gap $\Delta E$ of the detector, leading to conceptual challenges regarding the notion of a detector-independent temperature.
	Therefore, the circular Unruh effect cannot be connected to an acceleration-dependent, but detector-independent, temperature in the same sense as the usual Unruh effect.
	Instead, it should be understood in the broader context of acceleration-induced quantum transitions through non-inertial motion.
	
	So far, most discussions of the circular Unruh effect~\cite{Letaw1980, Letaw1981, Crispino2008, Bell1983, Hacyan1986, Bell1987, Kim1987, Levin1993, Davies1996, Unruh1998, Rosu2005, Lochan2020, Zhou2025, Zheng2025, Parry2025} have focused either on electric dipole interactions~\cite{Rosu2005, Lochan2020, Zhou2025, Audretsch1995, Arya2023}---often simplified via scalar-field approximations~\cite{Kim1987, Levin1993, Davies1996, Rosu2005, Zheng2025, Parry2025} that neglect both the spatial orientation of the atomic dipole and the full vector nature of the field---or on the coupling between electron spin and the electromagnetic vacuum~\cite{Bell1983, Bell1987, Unruh1998}.
	In contrast, we present a comprehensive framework that incorporates not only the electric but also the magnetic dipole transitions of the multilevel atom.
	By analyzing the spatial orientation of both matrix elements alongside the full vector structure of the field, we uniquely capture the distinct contributions and interplay of electric versus magnetic interactions in the circular Unruh effect.
	
	We model the atom as a three-level system consisting of a ground state $\ket{g}$ and two excited states $\ket{e}$ and $\ket{m}$, with $\ket{g}$ coupled either to $\ket{e}$ or to $\ket{m}$ via electric or magnetic dipole transition, respectively.
	The atomic center-of-mass motion is treated classically, i.\,e., not quantized, with the atom following a planar circular trajectory at a constant angular velocity $\alpha$.
	
	In the framework of this model, we obtain the transition rate of the circular Unruh effect, {\it i.e.}, the transition from $\ket{g}$ to $\ket{e}$ or to $\ket{m}$, as well as the rate of spontaneous emission, as the inverse process, i.\,e., the transition from $\ket{e}$ or $\ket{m}$ back to $\ket{g}$.
	It turns out that, for both electric and magnetic transitions, the spontaneous emission always dominates the circular Unruh transition, regardless of whether the electromagnetic field is given in free space or in a cylindrical cavity.
	To overcome this problem and hence to detect the circular Unruh effect, we propose a mitigation scheme that reduces the spontaneous emission rate by a factor of $1/2$, due to the constructive interference~\cite{Zhu1995}.
	In addition, to improve the feasibility of the circular Unruh effect, the coupling between the atom and the electromagnetic field needs to be increased.
	This can be achieved by integrating our setup into a cylindrical cavity and minimizing the mode volume, as the magnitude of the atom-light interaction is inversely proportional to the mode volume~\cite{Scully1997, CohenTannoudji1998, Schleich2001}.
	Taking the cavity as small as possible, while ensuring that it is still large enough to avoid any Casimir-Polder effects~\cite{Casimir1948, Lamoreaux2005, Buhmann2013}, a positive effective circular Unruh transition rate only occurs for the magnetic dipole coupling and a large angular velocity $\alpha$ compared to the corresponding transition frequency.
	According to our study and the available techniques to achieve large $\alpha$, the magnetic dipole transitions dominate when probing the circular Unruh effect.
	
	This article is structured as follows.
	In Sec.~\ref{sec: circular Unruh effect via electric and magnetic dipole transitions} we introduce our theoretical framework and present the analytical results for the rates of the circular Unruh transition (direct process) and spontaneous emission (inverse process) in free-field and cavity setups.
	Based on the mitigation scheme, discussed in Sec.~\ref{sec: Realization and feasibility}, we examine the efficiency of both processes depending on the type of atomic transition (electric versus magnetic) and the cavity size.
	Our conclusion in Sec.~\ref{Sec: Conclusion} is followed by two appendices.
	Appendix~\ref{app: transition rates} contains the detailed calculations of the transition rates in the free-space configuration, while in Appendix~\ref{app: minimal radius}, we derive a condition for the minimum radius of a cylindrical cavity required for an atom on a circular trajectory to detect the circular Unruh effect.
	
	\section{Circular Unruh effect via electric and magnetic dipole transitions}\label{sec: circular Unruh effect via electric and magnetic dipole transitions}
	An atom with multiple internal states~\cite{Scully1997, CohenTannoudji1998, Schleich2001, Steck2007} can encode signatures of its non-inertial center-of-mass motion into the transitions between these internal states.
	These transitions are induced by the interaction of the electric or magnetic dipole moments of the atom~\cite{Scully1997, CohenTannoudji1998, Schleich2001, Steck2007, Alden2014, AldenDiss2014, Schwartz2019, Schwartz2020b, Janson2022, Janson2024, Janson2025} with an external electromagnetic field.
	In an interaction picture with respect to the atom's internal degrees of freedom, the corresponding Hamiltonian~\cite{Scully1997, CohenTannoudji1998, Schleich2001, Steck2007, Alden2014, AldenDiss2014, Janson2022, Janson2025} is given by
	\begin{align}\label{eq: Hamiltonian magnetic dipole interaction}
		\hat{H}(t) = -\frac{e}{m} \hat{\vec{p}}(t) \cdot \hat{\vec{A}}(\vec{r},t) + \hat{\vec{\mu}}(t) \cdot \hat{\vec{B}}(\vec{r},t).
	\end{align}
	Here, $\hat{\vec{A}}(\vec{r},t)$ and $\hat{\vec{B}}(\vec{r},t) = \nabla \times \hat{\vec{A}}(\vec{r},t)$ are the quantized electromagnetic vector potential and the magnetic field, whereas $e$ and $m$ denote the charge and the (reduced) mass of the electron, respectively.
	Both $\hat{\vec{A}}(\vec{r},t)$ and $\hat{\vec{B}}(\vec{r},t)$ are evaluated at the classical center-of-mass trajectory $\vec{r} = \vec{r}_{\rm cl}(t)$.
	
	Moreover, for an atom modeled by a three-level system, the relative momentum and the magnetic dipole operators are given by \mbox{$\hat{\vec{p}}(t) = \ee^{\ii \omega_{eg}t}\vec{p}_{eg}\ket{e}\bra{g} + \ee^{-\ii \omega_{eg}t}\vec{p}_{eg}^{*}\ket{g}\bra{e}$} and \mbox{$\hat{\vec{\mu}}(t) = \ee^{\ii \omega_{mg}t} \vec{\mu}_{mg} \ket{m}\bra{g} + \ee^{-\ii \omega_{mg}t} \vec{\mu}_{mg}^{*} \ket{g}\bra{m}$}, where $\omega_{ig}$, with $i=e$ or $i=m$, is the transition frequency between the states $\ket{g}$ and $\ket{i}$.
	We consider only electric dipole transitions between the states $\ket{g}$ and $\ket{e}$, and only magnetic dipole transitions between the states $\ket{g}$ and $\ket{m}$.
	
	Note that beyond the rotating-wave approximation~\cite{Scully1997, CohenTannoudji1998, Steck2007}, the $\hat{\vec{p}}\cdot\hat{\vec{A}}$ coupling is not equivalent to the well-known $\hat{\vec{d}}\cdot\hat{\vec{E}}$ term, with $\hat{\vec{d}}$ and $\hat{\vec{E}}$ being the electric dipole operator and the electric field, respectively.
	The non-equivalence of these interaction terms originates from the counter-rotating terms, which are usually neglected and do not require the resonance condition $\omega\approx\omega_{ig}$ for the frequency $\omega$ of the electromagnetic field~\cite{Funai2019}, provided that the atom is truncated to an effective $n$-level system with finite $n$.
	Since the Unruh effect is exactly induced by these counter-rotating terms, we use the $\hat{\vec{p}}\cdot\hat{\vec{A}}$ coupling given in Eq.~\eqref{eq: Hamiltonian magnetic dipole interaction}.
	
	\begin{figure}[!htbp]
		\includegraphics[width=\columnwidth]{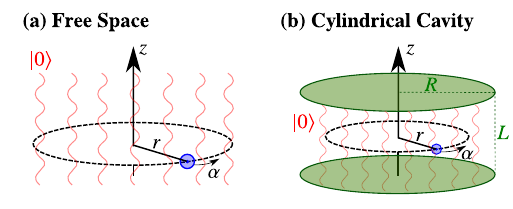}
		\caption{Two setups for detecting the circular Unruh effect using multilevel atoms. Panel \textbf{(a)} shows the center-of-mass motion of an atom (blue filled circle) confined to a circular trajectory of radius $r$ and angular velocity $\alpha$ in free space. The atom interacts with a background electromagnetic field (red waves) in the vacuum state $\ket{0}$. Panel \textbf{(b)} shows the center-of-mass motion of an atom confined to a circular trajectory of radius $r$ and angular velocity $\alpha$ inside a cavity (two green plates) of length $L$ and radius $R$. The atom interacts with a background electromagnetic field (red waves) in the vacuum state $\ket{0}$. To minimize the mode volume $V = \pi R^2 L$ and thereby to enhance the atom-light interaction, we consider a vanishingly small length $L \to 0$, while the radius is treated as infinitely large, $R \to \infty$, in order to satisfy the minimal radius condition found in Appendix~\ref{app: minimal radius}.}
		\label{fig:scheme}
	\end{figure} 
	
	For detecting the circular Unruh effect using multilevel atoms, we consider two setups, shown in Fig.~\ref{fig:scheme}.
	In the first one, denoted as free field and presented in Fig.~\ref{fig:scheme}(a), the atomic center-of-mass moves on a planar circular trajectory 
	\begin{align}
		\label{eq: circular classical trajectory}
		\vec{r}_{\rm cl}(t) = r \left[\cos(\alpha t), \sin(\alpha t), 0\right]^\text{T},
	\end{align}
	of radius $r$ with an angular velocity $\alpha$. 
	The confined atom interacts with an external free-space electromagnetic field prepared in the vacuum state $\ket{0}$, defined in the inertial laboratory frame.
	
	In the second setup, denoted as a cavity setup and presented in Fig.~\ref{fig:scheme}(b), the atomic center-of-mass follows the classical trajectory given by Eq. \eqref{eq: circular classical trajectory} inside a cylindrical cavity of length $L$ and radius $R$~\cite{Kakazu1996, Stroehle2024}.
	Again, the trapped atom interacts with an electromagnetic background field prepared in the vacuum state $\ket{0}$. 
	
	The main idea for using a cavity originates from the fact that the strength of the atom-light interaction scales with the inverse $1/V$ of the mode volume $V$.
	As a result, by minimizing $V$, the transition rate $\Gamma$ can be maximized. 
	However, in App.~\ref{app: minimal radius} we show that the cylindrical cavity with radius $R$ can be effectively described as two infinitely large plates for our purpose, i.\,e., $R \to \infty$ in Fig.~\ref{fig:scheme}(b), so that we can only minimize the mode volume $V = \pi R^2 L$ by minimizing the cavity length $L$.
	
	\subsection{Free-space electromagnetic field}\label{sec: free space}
	The vector potential of the electromagnetic field without boundary conditions in free space reads~\cite{Scully1997, CohenTannoudji1998, Schleich2001}
	\begin{align}\label{eq: vector potential free field}
		\hat{\vec{A}}(\vec{r},t)\!=\!\sum_{\lambda=1}^2\!\int\!\frac{\dd^3k}{(2\pi)^{3/2}} \left\{\vec{\mathcal{A}}_\lambda(\vec{k}) \hat{a}_\lambda(\vec{k})\ee^{\ii[\vec{k}\cdot\vec{r}(t) - \omega(k) t]} + \mathrm{h.c.}\right\}
	\end{align}
	with the amplitude 
	\begin{align}
		\vec{\mathcal{A}}_\lambda(\vec{k}) = \sqrt{\frac{\hbar}{2\epsilon_0\omega(k)}}\vec{e}_\lambda(\vec{k})
	\end{align}
	and the frequency $\omega(k) = c k$. 
	Here $\epsilon_0$ is the vacuum permittivity and the vectors $\vec{e}_\lambda(\vec{k})$, with $\lambda \in \{1,2\}$, represents the two possible polarizations. 
	In addition, we define the annihilation $\hat{a}_\lambda(\vec{k})$ and the creation $\hat{a}^\dagger_\lambda(\vec{k})$ operators, obeying the commutation relation $\left[\hat{a}_\lambda(\vec{k}),\hat{a}^\dagger_{\lambda^\prime}(\vec{k}^\prime)\right]=\delta_{\lambda\lambda^\prime}\delta(\vec{k}-\vec{k}^\prime)$ with the Kronecker symbol $\delta_{\lambda\lambda^\prime}$ and the Dirac delta distribution $\delta(\vec{k}-\vec{k}^\prime)$.
	Moreover, we define the electric and magnetic Rabi coupling strengths~\cite{Rabi1937}
	\begin{subequations}\label{eq: Rabi frequencies}
		\begin{align}
			\Omega_{\lambda}^{e}(\vec{k}) &= -\frac{e \vec{p}_{eg} \cdot \vec{\mathcal{A}}_\lambda(\vec{k})}{m\hbar} \equiv \vec{\Omega}^{e}(k)\cdot\vec{e}_\lambda(\vec{k}),\\
			\Omega^{m}_\lambda(\vec{k}) &= \frac{\ii \vec{\mu}_{mg} \cdot \left[\vec{k} \times \vec{\mathcal{A}}_\lambda(\vec{k})\right]}{\hbar} \equiv \vec{\Omega}^{m}(k)\cdot \frac{\vec{k}\times \vec{e}_\lambda(\vec{k})}{\vert\vec{k}\vert},
		\end{align}
	\end{subequations}
	quantifying the efficiency of the electric and magnetic transitions, respectively. 
	In the following, we restrict our treatment to Rabi frequencies that are constant in time, which can be achieved by aligning the dipole matrix elements along a certain quantization axis with a constant magnetic field.
	
	To detect the Unruh effect, the Unruh transition rate $\Gamma^i_+$ has to supersede the corresponding spontaneous emission rate $\Gamma^i_-$. 
	In the case of the Unruh effect, the atom is initially prepared in its ground state $\ket{g}$ and then transitions to the excited state $\ket{i}$.
	In the case of the spontaneous emission, being the inverse process with respect to the Unruh transition, the atom decays from the excited state $\ket{i}$ into the ground state $\ket{g}$.
	In both cases, the electromagnetic field is initially in its vacuum state $\ket{0}$, with $\hat{a}_\lambda(\vec{k})\ket{0}=0$, while its final state is the state of a photon of arbitrary polarization and frequency. 
	Thus, the initial state of the whole atom-field system is given by $\ket{\Psi}^\text{(U)}_{\mathrm{in}} = \ket{g} \otimes \ket{0}$ for the Unruh effect, and $\ket{\Psi^{i}}^{\text{(S)}}_{\mathrm{in}} = \ket{i} \otimes \ket{0}$ for the spontaneous emission process.
	
	By using the first-order Dyson series~\cite{Dyson1949}, we derive in Appendix~\ref{app: transition rates} the analytical equations for $\Gamma^i_+$, defined as the probability $P^i_+(T)$ per interaction time $T$ for the atom to be found in the excited state $\ket{i}$ in the long-time limit $T \to \infty$.
	Similarly, we derive the relation for $\Gamma^i_-$.
	They read
	\begin{align}\label{eq: Transition rate (free space)}
		\begin{split}
			\Gamma^{i}_\pm 
			&= \frac{1}{4\pi c}\!\!\!\!\sum_{n = \lceil\pm\frac{\omega_{ig}}{\alpha}\rceil}^\infty\!\!\!\!\left[k^{i,(\pm)}_{n} \vert\vec{\Omega}^{i}(k^{i,(\pm)}_{n})\vert\right]^2\!\int\limits_{-1}^{1}\!\!\!\dd x J_n^2\left(r k^{i,(\pm)}_{n} \sqrt{1-x^2}\right)\\
			&\times\left[\sin^2\left(\theta_{\vec{\Omega}^i}\right) \left(1 + x^2\right) + 2\cos^2\left(\theta_{\vec{\Omega}^i}\right) \left(1 - x^2\right)\right]\\
			&\equiv
			\begin{cases}
				\cfrac{e^2 \vert\vec{p}_{eg}\vert^2 \omega_{eg}}{8\pi \hbar \epsilon_0 m^2 c^3}\, \tilde{\Gamma}^{(1)}_\pm(\tilde{\alpha}_e,\tilde{v}), & \text{for } i=e\\
				\\
				\cfrac{\vert\vec{\mu}_{mg}\vert^2 \omega_{mg}^3}{8\pi \hbar \epsilon_0 c^5}\, \tilde{\Gamma}^{(3)}_\pm(\tilde{\alpha}_m,\tilde{v}), & \text{for } i=m,
			\end{cases}
		\end{split}
	\end{align}
	where $\theta_{\vec{\Omega}^i}$ is the angle between the Rabi vector $\vec{\Omega}^i$ defined in Eq.~\eqref{eq: Rabi frequencies} and the $z$-axis shown in Fig.~\ref{fig:scheme}, $r$ is the radius of the circle, and $\lceil x\rceil$ denotes the least integer greater than or equal to $x$.
	Here, the wave vector
	\begin{align}
		k^{i,(\pm)}_{n} =\frac{n\alpha \mp \omega_{ig}}{c} = \frac{\omega_{ig}}{c}\left(n\tilde{\alpha}_i\mp 1\right) = \frac{\omega_{ig}}{c}\,\tilde{k}_n^{i,(\pm)}
	\end{align}
	is rewritten in terms of the dimensionless wave vector $\tilde{k}_n^{i,(\pm)}$ and the dimensionless angular velocity \mbox{$\tilde{\alpha}_i = \alpha / \omega_{ig}$}.
	
	In addition, we have introduced the dimensionless transition rates
	\begin{align}\label{eq: Dimensionless transition rate (free space)}
		\begin{split}
			\tilde{\Gamma}^{(j)}_\pm(\tilde{\alpha}_i,\tilde{v}) &=\sum_{n=\lceil\pm\tilde{\alpha}_i^{-1}\rceil}^\infty\left[\tilde{k}_n^{i,(\pm)}\right]^j\int\limits_{-1}^{1}\mathrm{d}x J_n^2\left(\frac{\tilde{v}}{\tilde{\alpha}_i}\tilde{k}_n^{i,(\pm)} \sqrt{1-x^2}\right) \\
			&\times \left[\sin^2\left(\theta_{\vec{\Omega}^i}\right) \left(1 + x^2\right) + 2 \cos^2\left(\theta_{\vec{\Omega}^i}\right) \left(1 - x^2\right)\right]
		\end{split}
	\end{align}
	with $\tilde{v} = v/c$.
	
	Since we consider only non-relativistic velocities $v \ll c$, we expand Eq.~\eqref{eq: Dimensionless transition rate (free space)} in terms of $\tilde{v}$ up to the first non-trivial order, obtaining
	\begin{subequations}\label{eq: Dimensionless transition rates (free space) expanded}
		\begin{align}
			\begin{split}
				\tilde{\Gamma}^{(j)}_+ (\tilde{\alpha}_i,\tilde{v})&\approx
				\frac{4(n_\text{min}^i + 1) [4+3 n_\text{min}^i+n_\text{min}^i \cos(2\theta_{\vec{\Omega}^i})]}{(2n_\text{min}^i+3)!}\\
				&\times (n_\text{min}^i \tilde{\alpha}_i - 1)^{j+2 n_\text{min}^i}\left(\frac{\tilde{v}}{\tilde\alpha_i}\right)^{2 n_\text{min}^i} 
			\end{split}\\
			\begin{split}
				\tilde{\Gamma}^{(j)}_- (\tilde{\alpha}_i,\tilde{v})&\approx \frac{8}{3} + \frac{7 + \cos(2\theta_{\vec{\Omega}^i})}{15}\left(\frac{\tilde{v}}{\tilde{\alpha}_i}\right)^2\\
				&\times
				\begin{cases}
					(\tilde{\alpha}_i + 1)^{2+j} - (\tilde{\alpha}_i - 1)^{2+j} - 2, &\text{for } \tilde{\alpha}_i < 1\\
					(\tilde{\alpha}_i + 1)^{2+j} - 2, & \text{for } \tilde{\alpha}_i > 1
				\end{cases}
			\end{split}
		\end{align}
	\end{subequations}
	with $j \in \{1,3\}$ and $n_\text{min}^i = \lceil\tilde{\alpha}_i^{-1}\rceil$.
	
	\begin{figure*}[!htbp]
		\begin{subfigure}[b]{\textwidth}
			\includegraphics{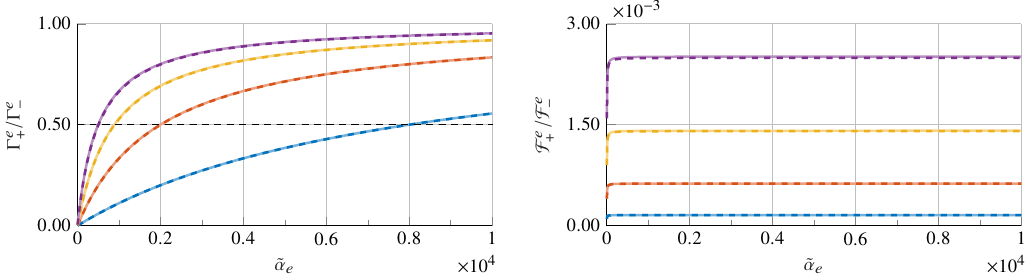}
			\caption{Circular Unruh transition rate versus spontaneous emission rate via electric dipole transitions.}
			\label{fig:gamma_Unruh_vs_Spont electric}
		\end{subfigure}
		
		\vspace{0.5cm} 
		
		\begin{subfigure}[b]{\textwidth}
			\includegraphics{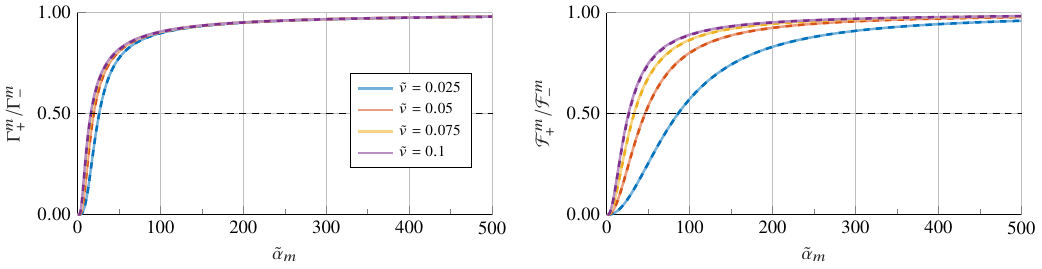}
			\caption{Circular Unruh transition rate versus spontaneous emission rate via magnetic dipole transitions.}
			\label{fig:gamma_Unruh_vs_Spont magnetic}
		\end{subfigure}
		\caption{Ratio of the circular Unruh transition rate and the spontaneous emission rate via (a) electric and (b) magnetic dipole transitions in an atom moving on a circle with radius $r$ and angular velocity $\alpha$, see Fig. \ref{fig:scheme}. The left and right panels show the results for the free-space ($\Gamma^m_+ / \Gamma^m_-$), and a cavity setup ($\mathcal{F}^m_+ / \mathcal{F}^m_-$) for distinct orbital velocities $\tilde{v}= \alpha r / c$.
			The ratios are plotted against the normalized angular velocities $\tilde\alpha_i = \alpha / \omega_{ig}$, with $\omega_{ig}$ being the transition frequency between the states $\ket{g}$ and $\ket{i}$, where $i\in\{e,m\}$. 
			Solid lines are obtained via numerical simulations, while the dashed lines are given by Eqs.~\eqref{eq: Transition rate (free space)} and \eqref{eq: Dimensionless transition rates (free space) expanded}, or Eqs.~\eqref{eq: Transition rates (two plates)} and \eqref{eq: Dimensionless transition rates (cavity) expanded}.
			To achieve a positive effective Unruh transition rate $\Gamma^i_\text{eff} = \Gamma^i_+ - \Gamma^i_-/2$ or $\mathcal{F}^i_\text{eff} = \mathcal{F}^i_+ - \mathcal{F}^i_-/2$, both ratios $\Gamma^{i}_{+} / \Gamma^{i}_{-}$ and $\mathcal{F}^{i}_{+} / \mathcal{F}^{i}_{-}$ must exceed $1/2$ (black dashed line).}
		\label{fig:gamma_Unruh_vs_Spont}
	\end{figure*}
	
	Both transition rates, $ \tilde{\Gamma}^{(j)}_{+}$ and $ \tilde{\Gamma}^{(j)}_{-}$, achieve their maximum values for $\theta_{\vec{\Omega}^i} = 0$ or $\theta_{\vec{\Omega}^i} = \pi$ and their minimal values for $\theta_{\vec{\Omega}^i} = \pi/2$.
	Moreover, the transition rates $\tilde{\Gamma}^{(j)}_{+}$ and $\tilde{\Gamma}^{(j)}_{-}$ depend polynomially on $\tilde{\alpha}_i$ and grow with increasing $\tilde{\alpha}_i$.
	However, for all $\tilde{\alpha}_i\geq 0$, one can prove that $\tilde{\Gamma}^{(j)}_+ < \tilde{\Gamma}^{(j)}_-$, as shown in the left subfigures of Fig. \ref{fig:gamma_Unruh_vs_Spont}.
	Only for $\tilde{\alpha}_i\to\infty$ we find $\tilde{\Gamma}^{(j)}_{+} / \tilde{\Gamma}^{(j)}_{-} \to 1$. 
	
	\subsection{Cylindrical cavity}\label{sec: cylindrical cavity}
	As already discussed in the previous section, the coupling between the atom and the electromagnetic field is proportional to the inverse mode volume $1/V$.
	At the same time, minimizing the mode volume inherently decreases the mode density.
	Consequently, maximizing the transition rate involves an interplay between minimizing the mode volume and avoiding an excessive reduction in the density of modes.
	The symmetry of the atomic center-of-mass motion in this setting naturally suggests a cylindrical cavity geometry~\cite{Kakazu1996, Stroehle2024} of radius $R$ and length $L$.
	This geometry has previously been considered~\cite{Levin1993, Davies1996, Zheng2025} in the context of scalar models, where the minimum radius requirement $\alpha R > c$ defines a threshold necessary to satisfy the resonance condition for the counter-rotating coupling.
	However, by taking into account the vector and spatial structure of the electromagnetic field in the cavity and by considering atoms coupled to the electromagnetic field through a stationary dipole matrix element (in the laboratory frame), we obtain in Appendix~\ref{app: minimal radius} the lower bound $\alpha R > c$ for dipole matrix elements along the $z$-direction, and 
	\begin{align}\label{eq: condition minimal radius}
		\alpha R \gtrsim 0.92\,c
	\end{align}
	for dipole matrix elements in the $xy$-plane.
	This is a consequence of the stationary matrix elements rotating with respect to the field modes.
	
	Moreover, working in the non-relativistic regime defined by \mbox{$v/c = \alpha r / c \ll 1$} yields the condition that the radius $r$ of the atom's circular center-of-mass motion has to be much smaller than the cavity radius $R$.
	This condition, in turn, implies that the cavity is effectively unbounded in the radial direction, such that $R \to \infty$ can be effectively assumed.
	Therefore, it is justified to reduce the treatment to the limiting case \mbox{$L \to 0$}, corresponding to two parallel, infinitely large plates separated by a vanishingly small distance.
	
	\subsubsection{Infinitely large plates}\label{sec: infinitely large plates}
	The electromagnetic field in between two parallel, infinitely large plates can be derived from the free field, Eq.~\eqref{eq: vector potential free field}, by replacing~\cite{Scully1997, CohenTannoudji1998, Schleich2001, Steck2007}
	\begin{align}
		\int\frac{\dd^3k}{(2\pi)^{3/2}} \rightarrow \sum_{\ell=0}^\infty \int\frac{\dd^2k_\perp}{2\pi\sqrt{L}},
	\end{align}
	and discretizing the $z$-component of the wave-vector \mbox{$\vec{k} = \left(\vec{k}_\perp, \ell\pi/L\right)^\text{T}$}, with $\ell \in \mathbb{N}_0$.
	
	Following the same approach as in the free-field case presented in Appendix~\ref{app: transition rates} yields the analytical expressions
	\begin{align}\label{eq: Transition rates (two plates)}
		\begin{split}
			\mathcal{F}^i_\pm &= \frac{1}{4 c L} \left(1+\cos^2\theta_{\vec{\Omega}^i}\right) \\
			&\times \sum_{n=\lceil\pm\frac{\omega_{ig}}{\alpha}\rceil}^\infty k^{i,(\pm)}_{\perp,(n,\ell=0)} \left\vert \vec{\Omega}^i(k^{i,(\pm)}_{\perp(n,\ell=0)}) \right\vert^2 J_{n}^2\left[r k^{i,(\pm)}_{\perp,(n,\ell=0)}\right]\\
			&\equiv
			\begin{cases}
				\cfrac{e^2 \vert\vec{p}_{eg}\vert^2}{4\hbar\epsilon_0 m^2 c^2 L}\, \tilde{\mathcal{F}}^{(0)}_{\pm}(\tilde{\alpha}_e,\tilde{v}), & \text{for } i = e\\
				\\
				\cfrac{\vert\vec{\mu}_{mg}\vert^2 \omega_{mg}^2}{4\hbar\epsilon_0 c^4 L}\, \tilde{\mathcal{F}}^{(2)}_{\pm}(\tilde{\alpha}_m,\tilde{v}), & \text{for } i = m
			\end{cases}
		\end{split}
	\end{align}
	for the rate $\mathcal{F}^i_+$ of the circular Unruh transition and the rate $\mathcal{F}^i_-$ of the inverse process, i.\,e., the spontaneous emission, in the case of the cylindrical cavity. 
	Here, the amplitude of the transverse component of the wave vector
	\begin{align}
		k^{i,(\pm)}_{\perp,(n,\ell)} = \sqrt{\left(\frac{n\alpha \mp \omega_{ig}}{c}\right)^2 - \left(\frac{\ell\pi}{L}\right)^2}
	\end{align}
	has to be a real quantity, implying \mbox{$\left(n\alpha \mp \omega_{ig}\right) / c \geq \left(\ell\pi\right) / L$}. 
	In the limit $L \to 0$, we have \mbox{$n_{\pm,\text{min}}(\ell>0) \gg n_{\pm,\text{min}}(\ell=0)$}.
	Therefore, the modes with $\ell = 0$ are dominant and all modes with $\ell>0$ can be neglected, leading us to $k_z = 0$ and $k = k_\perp$.
	Hence, the initially three-dimensional problem is reduced to an effective two-dimensional setting.
	
	In addition, we have introduced in Eq.~\eqref{eq: Transition rates (two plates)} the dimensionless transition rates
	\begin{align}\label{eq: Dimensionless transition rates (two plates)}
		\tilde{\mathcal{F}}^{(j^\prime)}_\pm(\tilde{\alpha}_i,\tilde{v}) = (1 + \cos^2\theta_{\vec{\Omega}^i}) \sum_{n=\lceil\pm\tilde{\alpha}_i^{-1}\rceil}^\infty \left[\tilde{k}_n^{i,(\pm)}\right]^{j^\prime} J_n^2\left[\frac{\tilde{v}}{\tilde{\alpha}_i}\tilde{k}_n^{i,(\pm)}\right],
	\end{align}
	where $j^\prime\in \{0,2\}$ is reduced by one compared to the free-field setup of Sec.~\ref{sec: free space}, since the cavity effectively reduces the $k$-space dimensionality from three to two dimensions, and we have again defined $n_\text{min}^i = \lceil\tilde\alpha_i^{-1}\rceil$.
	As in section \ref{sec: free space}, expanding the result Eq. \eqref{eq: Dimensionless transition rates (two plates)} in terms of $\tilde{v}$ up to the first non-trivial order yields
	\begin{subequations}\label{eq: Dimensionless transition rates (cavity) expanded}
		\begin{align}
			\tilde{\mathcal{F}}^{(j^\prime)}_+\!\!&\approx\!
			(1 + \cos^2\theta_{\vec{\Omega}^i})\frac{(n_\text{min}^i \tilde{\alpha}_i - 1)^{j^\prime+2n_\text{min}^i}}{4[(n_\text{min}^i)!]^2} \left(\frac{\tilde{v}}{\tilde\alpha_i}\right)^{2n_\text{min}^i}\\
			\tilde{\mathcal{F}}^{(j^\prime)}_-\!\!&\approx\!(1+\cos^2\theta_{\vec{\Omega}^i}) \nonumber \\
			&\times
			\begin{cases}
				\!1\!+\!\cfrac{\tilde{v}^2}{4\tilde{\alpha}_i^2} \left[(\tilde{\alpha}_i + 1)^{2+j^\prime} - (\tilde{\alpha}_i - 1)^{2+j^\prime} - 2\right], & \text{for } \tilde{\alpha}_i\!<\!1\\
				\\
				\!1\!+\!\cfrac{\tilde{v}^2}{4\tilde{\alpha}_i^2} \left[(\tilde{\alpha}_i + 1)^{2+j^\prime} - 2\right], & \text{for } \tilde{\alpha}_i\!>\!1.
			\end{cases}
		\end{align}
	\end{subequations}
	
	As in the previous case of section \ref{sec: free space}, the transition rates are maximal for $\theta_{\vec{\Omega}^i} = 0$ or $\theta_{\vec{\Omega}^i} = \pi$ and minimal for $\theta_{\vec{\Omega}^i} = \pi/2$.
	Moreover, the Unruh transition rate is always smaller than the spontaneous emission rate, i.\,e., $\tilde{\mathcal{F}}^{(j^\prime)}_+ < \tilde{\mathcal{F}}^{(j^\prime)}_-$.
	Furthermore, in the limit of large angular velocities, we obtain
	\begin{subequations}\label{eq: limit alpha to infty (plates)}
		\begin{align}
			\lim_{\tilde\alpha_e\to\infty} \frac{\tilde{\mathcal{F}}^{(0)}_+}{\tilde{\mathcal{F}}^{(0)}_-} &= \frac{\tilde{v}^2}{4 + \tilde{v}^2} \leq \frac{1}{5},\\
			\lim_{\tilde\alpha_m\to\infty} \frac{\tilde{\mathcal{F}}^{(2)}_+}{\tilde{\mathcal{F}}^{(2)}_-} &= 1.
		\end{align}
	\end{subequations}
	
	This result implies that in the cylindrical cavity setup and for the electric dipole transitions, $i=e$ or $j^\prime=0$, it is not possible to achieve positive effective Unruh transition rates, even if we suppress the spontaneous emission by the mitigation scheme presented in Sec.~\ref{sec: Mitigation Scheme}.
	This assessment is a direct consequence of the reduction from three to effectively two dimensions in the case of the cylindrical cavity, and the corresponding decrease in the density of modes.
	Indeed, in free space, the transition rates $\Gamma^e_\pm$ via electric dipole transitions, given by Eq. \eqref{eq: Transition rate (free space)} with $i=e$ and Eq. \eqref{eq: Dimensionless transition rates (free space) expanded} with $j=1$,  are first-order polynomials in $\alpha$ as $\alpha \to \infty$.
	However, for the cavity setup, the corresponding transition rates $\mathcal{F}^e_\pm$, given by Eq. \eqref{eq: Transition rates (two plates)} with $i=e$ and Eq. \eqref{eq: Dimensionless transition rates (cavity) expanded} with $j^\prime=0$,  are zeroth-order polynomials in $\alpha$ as $\alpha \to \infty$.
	
	An effective temperature defined via the detailed-balance condition discussed in Sec.~\ref{sec: Introduction} would not only depend on the energy gap $\hbar \omega_{ig}$, but also on the type of interaction (electric or magnetic dipole transitions), in contrast to the usual Unruh effect through linear accelerations.
	This assessment further highlights the conceptual limitations of defining a temperature in circular setups.
	
	\section{Realization \& feasibility}\label{sec: Realization and feasibility}
	As discussed before, the Unruh transition competes with the spontaneous emission.
	According to the previous section, it is not possible for the circular Unruh transition rate to exceed the spontaneous emission rate in the free-field setup and a cavity setup with (infinitely) large radius $R$.
	In this section we propose a suitable mitigation scheme to reduce the spontaneous emission rate.
	
	\subsection{Mitigation Scheme}\label{sec: Mitigation Scheme}
\begin{figure}[t]
		\centering
		\includegraphics[width=\columnwidth]{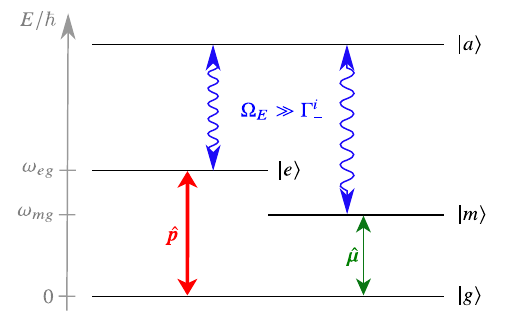}
		\caption{Scheme of the relevant internal-energy levels in a multilevel atom with ground state $\ket{g}$ and two excited states $\ket{e}$ and $\ket{m}$, being separated by the energy spacing $\Delta E = \hbar \omega_{ig}$ (with $i\in \{e,m\}$) and coupled via electric ($\hat{\vec{p}}$, red) or magnetic ($\hat{\vec{\mu}}$, green) dipole transitions, together with an ancilla state $\ket{a}$. An external laser field (blue wiggled lines) couples the ancilla state $\ket{a}$ to either $\ket{e}$ or $\ket{m}$ in order to reduce the spontaneous emission rate~\cite{Zhu1995}, depending on which of these excited states provides the dominant circular Unruh coupling to the ground state $\ket{g}$. The (electric) Rabi frequency $\Omega_E$ is assumed to be large enough compared to the spontaneous emission rate $\Gamma^i_-$ of the state $\ket{i}$.}
		\label{fig: level structure}
	\end{figure}

	In order to find a positive effective Unruh transition rate, which is defined as the difference between the Unruh transition rate and the spontaneous emission rate, the former must exceed the latter.
	Note that this scheme is still sensitive to the circular Unruh effect by measuring the ground state instead of the excited-state population.
	To achieve this goal, we consider a laser field that couples the excited state $\ket{i}$ to a third (ancilla) state $\ket{a}$ via electric dipole transitions with a Rabi frequency $\Omega_E$ that is large enough compared to the spontaneous emission rate $\Gamma^{i}_{-}$ of the state $\ket{i}$, as illustrated in Fig.~\ref{fig: level structure}.
	Additionally, it is assumed that no dipole transitions are allowed between the ancilla state $\ket{a}$ and the ground state $\ket{g}$.
	Under these conditions, the spontaneous emission rates are reduced by a factor of $1/2$~\cite{Zhu1995}.
	As a result, using Eqs.~\eqref{eq: Transition rate (free space)} and~\eqref{eq: Transition rates (two plates)}, the effective rates of the circular Unruh transition read
	\begin{subequations}
		\begin{align}
			\Gamma^i_\text{eff} &= \Gamma^i_+ - \frac{1}{2}\Gamma^i_- = \begin{cases}\label{eq: effective transition rates free field}
				\cfrac{e^2 \vert\vec{p}_{eg}\vert^2 \omega_{eg}}{8\pi \hbar \epsilon_0 m^2 c^3}\, \tilde\Gamma^{(1)}_\text{eff}, & \text{for } i = e\\
				\\
				\cfrac{\vert\vec{\mu}_{mg}\vert^2 \omega_{mg}^3}{8\pi \hbar \epsilon_0 c^5}\, \tilde\Gamma^{(3)}_\text{eff}, & \text{for } i = m
			\end{cases}\\
			\mathcal{F}^i_\text{eff} &= \mathcal{F}^i_+ - \frac{1}{2}\mathcal{F}^i_- = 
			\begin{cases}\label{eq: effective transition rates cavity}
				\cfrac{e^2 \vert\vec{p}_{eg}\vert^2}{4\hbar\epsilon_0 m^2 c^2 L}\, \tilde{\mathcal{F}}^{(0)}_\text{eff}, & \text{for } i = e\\
				\\
				\cfrac{\vert\vec{\mu}_{mg}\vert^2 \omega_{mg}^2}{4\hbar\epsilon_0 c^4 L}\, \tilde{\mathcal{F}}^{(2)}_\text{eff}, & \text{for } i = m
			\end{cases}
		\end{align}
	\end{subequations}
	with $\tilde\Gamma^{(j)}_\text{eff} = \tilde{\Gamma}^{(j)}_+ - \tilde{\Gamma}^{(j)}_-/2$ and $\tilde{\mathcal{F}}^{(j^\prime)}_\text{eff} = \tilde{\mathcal{F}}^{(j^\prime)}_+ - \tilde{\mathcal{F}}^{(j^\prime)}_-/2$.
	
	In terms of the dimensionless transition rates, the condition for a positive effective Unruh transition rate is given by $\tilde{\Gamma}^{(j)}_+/\tilde{\Gamma}^{(j)}_- > 1/2$, or $\tilde{\mathcal{F}}^{(j^\prime)}_+ / \tilde{\mathcal{F}}^{(j^\prime)}_- > 1/2$.
	According to Eq.~\eqref{eq: limit alpha to infty (plates)} and Fig.~\ref{fig:gamma_Unruh_vs_Spont}, this requirement can be achieved for $\tilde{\alpha}_i \gg 1$ for both electric and magnetic dipole transitions in free space, but only for magnetic dipole transitions in the cavity setup with two parallel infinitely large plates and small distance $L$.
	By using Eqs.~\eqref{eq: Dimensionless transition rates (free space) expanded} and \eqref{eq: Dimensionless transition rates (cavity) expanded}, we arrive at the expressions  
	\begin{subequations}\label{eq: dimensionless effective transition rates free field}
		\begin{align}
			\tilde\Gamma^{(1)}_\text{eff}(\tilde\alpha_e) &= -\frac{4}{3} + \left[7 + \cos(2\theta_{\vec{\Omega}^e})\right] \left(\frac{\tilde\alpha_e}{30} - \frac{3}{10}\right) \tilde{v}^2\\
			\tilde\Gamma^{(3)}_\text{eff}(\tilde\alpha_m) &= -\frac{4}{3} + \left[7 + \cos(2\theta_{\vec{\Omega}^m})\right] \left(\frac{\tilde\alpha_m^3}{30} - \frac{\tilde\alpha_m^2}{2} + \frac{\tilde\alpha_m}{3} - 1\right) \tilde{v}^2,
		\end{align}
	\end{subequations}
	in the case of free space, and
	\begin{subequations}\label{eq: dimensionless effective transition rates cavity}
		\begin{align}
			\tilde{\mathcal{F}}^{(0)}_\text{eff} (\tilde\alpha_e) &= 0\\
			\tilde{\mathcal{F}}^{(2)}_\text{eff} (\tilde\alpha_m) &= \frac{1 + \cos^2\theta_{\vec{\Omega}^m}}{8} \left[-4 + (\tilde\alpha_m^2 - 12 \tilde\alpha_m + 6) \tilde{v}^2\right],
		\end{align}
	\end{subequations}
	in the case of the cavity setup for the dimensionless effective rates, as $\tilde{\alpha}_i \rightarrow \infty$. 
	It is important to note that $\tilde{\mathcal{F}}^{(0)}_\text{eff} (\tilde\alpha_e) = 0$ because $\mathcal{F}^{e}_{+} \ll \mathcal{F}^{e}_{-}$ for all $\tilde\alpha_e$, as shown in Fig.~\ref{fig:gamma_Unruh_vs_Spont electric}.
	
	\subsection{Electric versus magnetic dipole transitions}\label{sec: electric vs magnetic dipoles}
	In both setups, the polynomial power in $\tilde\alpha_m$ is two orders higher than the one of the polynomial in $\tilde\alpha_e$.
	In the cavity setup we immediately observe that the dominant coupling is given by the magnetic dipole interaction.
	To compare the effective transition rates for electric and magnetic dipole transitions in the case of free fields, we define the ratio of the electric and the magnetic dipole transition frequencies $\chi=\omega_{eg}/\omega_{mg}$, leading to the relation $\tilde\alpha_m = \chi \tilde\alpha_e$, and obtain the estimation
	\begin{align}\label{eq: electric vs magnetic}
		\begin{split}
			\frac{\Gamma^e_\text{eff}}{\Gamma^m_\text{eff}} &=
			\left(\frac{\vert\vec{d}_{eg}\vert}{\vert\vec{\mu}_{mg}\vert / c}\right)^2 \chi^3 \frac{\tilde\Gamma^{(1)}_\text{eff}(\tilde\alpha_e,\tilde{v})}{\tilde\Gamma^{(3)}_\text{eff}(\chi \tilde\alpha_e,\tilde{v})}\\
			&\lesssim \left(\frac{\vert\vec{d}_{eg}\vert}{\vert\vec{\mu}_{mg}\vert/c}\right)^2 \frac{\left[7 + \cos(2\theta_{\vec{\Omega}^e})\right]^3}{7 + \cos(2\theta_{\vec{\Omega}^m})} \frac{\tilde{v}^4}{10800} + \mathcal{O}\left(\frac{\tilde{v}^6}{\chi}\right).
		\end{split}
	\end{align}
	Here, we have used Eqs.~\eqref{eq: effective transition rates free field} and \eqref{eq: dimensionless effective transition rates free field}, keeping only the leading terms with respect to small $\tilde{v}$ and large $\chi$.
	Additionally, we have expressed $e\vert\vec{p}_{eg}\vert/m = \omega_{eg} \vert\vec{d}_{eg}\vert$ in terms of the well-known electric dipole matrix elements $\vec{d}_{eg} = \bra{e}\hat{\vec{d}}\ket{g}$. 
	
	\begin{figure}[t]
		\includegraphics{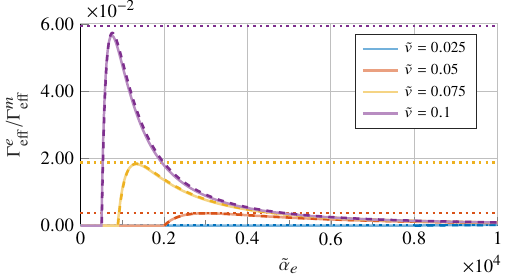}
		\caption{Ratio $\Gamma^e_\text{eff}/\Gamma^m_\text{eff}$ of the effective rates of the circular Unruh transition via electric $\Gamma^e_\text{eff}$, and magnetic $\Gamma^m_\text{eff}$ dipole transition as the function of the normalized angular velocity $\tilde\alpha_e = \alpha / \omega_{eg}$, for the four distinct scaled orbital velocities $\tilde{v} \in \{0.025, 0.05, 0.075, 0.1\}$.
			Here, we set $\chi = \omega_{eg}/\omega_{mg} = 10^3$, $\vert\vec{d}_{eg}\vert = e a_0$, $\vert\vec{\mu}_{mg}\vert = \mu_B$, $\theta_{\vec{\Omega}^e} = 0$, and $\theta_{\vec{\Omega}^m} = \pi/2$.
			Solid lines are obtained using full numerical calculation, while the dashed lines are given by Eqs.~\eqref{eq: effective transition rates free field} and \eqref{eq: dimensionless effective transition rates free field}.
			The dotted lines mark the corresponding maximum values given by Eq.~\eqref{eq: electric vs magnetic}.}
		\label{fig: Gamma_el_vs_mag}
	\end{figure}
	
	In the leading order, the estimation for the maximum of this ratio is independent of $\chi$.
	The ratio $\Gamma^e_\text{eff}/\Gamma^m_\text{eff}$ is plotted against $\tilde{\alpha}_e = \alpha / \omega_{eg}$ in Fig. \ref{fig: Gamma_el_vs_mag}.
	Consequently, the magnetic dipole transitions are the most suitable for observing the circular Unruh transition. 
	
	According to Eq.~\eqref{eq: dimensionless effective transition rates free field}, the effective rates for both electric and magnetic transitions grow as $\tilde\alpha_i$ increases.
	Therefore, minimizing the transition frequency $\omega_{ig}$ is highly beneficial.
	The lowest electric dipole transition frequencies are typically between internal states separated by the Lamb shift, {\it e.\,g.}, $\omega_{eg}/2\pi \approx 1\,\mathrm{GHz}$ for the $2s_{1/2} \to 2p_{1/2}$ transition~\cite{Lamb1947}.
	In contrast, the hyperfine splitting for states coupled by magnetic dipole transitions often falls in the $\mathrm{MHz}$ range~\cite{Reich1956}.
	
	Another strategy to minimize transition frequencies is the utilization of Rydberg atoms~\cite{Gallagher1994}.
	Since the energy spacing scales with $1/n^3$~\cite{Gallagher1994}, electric transition frequencies for high principal quantum numbers $n$ reach the $\mathrm{MHz}$ range, while magnetic dipole transition frequencies can drop in the $\mathrm{kHz}$ range~\cite{Cardman2022}.
	Thus, the ratio between the lowest available electric and magnetic dipole transition frequencies is approximately $\chi \approx 10^{3}$.
	However, when using Rydberg atoms, one has to account for cascade effects, where the atom decays into states lower than the designated ground state $\ket{g}$~\cite{Gallagher1994}.
	While additional laser fields can be used to repump the atoms back into the $\ket{g}$ state, the complexity of this setup increases with $n$, as a higher quantum number necessitates a greater number of repumping lasers.
	
	\subsection{Free-field versus cavity setup}\label{sec: free-field versus cavity setup}
\begin{figure}[!htbp]
		\includegraphics{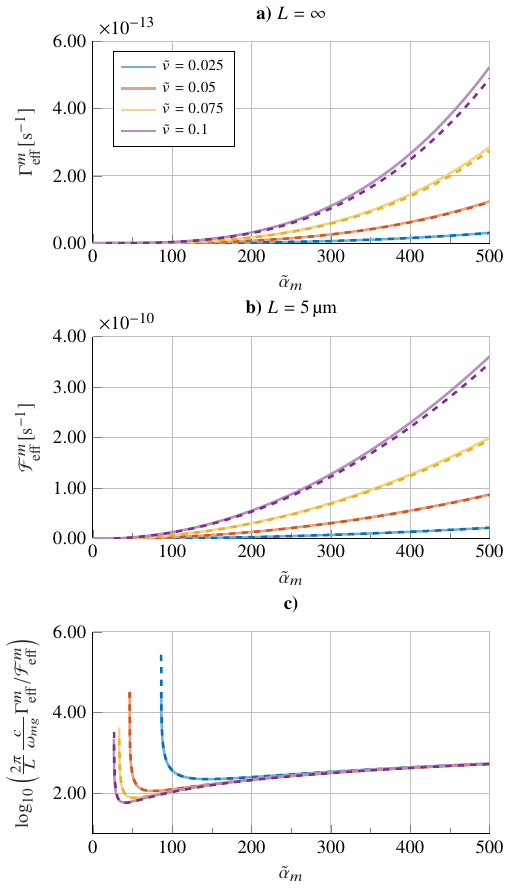}
		\caption{The effective circular Unruh transition rates via magnetic dipole transitions in a (a) free-space and (b) cavity setup with $L=5\,{\rm \mu m}$ plotted against the normalized angular velocity $\tilde\alpha_m = \alpha / \omega_{mg}$ for four distinct scaled orbital velocities $\tilde{v} \in \{0.025, 0.05, 0.075, 0.1\}$.
			Panel (c) shows the logarithm of the ratio $\tilde\Gamma^{(3)}_\text{eff}/\tilde{\mathcal{F}}^{(2)}_\text{eff}$ of the dimensionless effective transition rates.
			Solid lines are obtained via full numerical calculations, while the dashed lines are given by Eqs.~\eqref{eq: effective transition rates free field}, \eqref{eq: effective transition rates cavity}, \eqref{eq: dimensionless effective transition rates free field} and \eqref{eq: dimensionless effective transition rates cavity}.
			For all plots, we have used $\vert\vec{\mu}_{mg}\vert = \mu_B$ and $\omega_{mg} = 1\,\mathrm{GHz}$. }
		\label{fig: Gamma_mag_eff}
	\end{figure}

	In the previous section we have shown that the relevant coupling for measuring the circular Unruh effect is given by the magnetic dipole interaction. 
	In this section, our focus relies on these transitions.
	The corresponding effective rates $\Gamma^m_\text{eff}$ from Eq.~\eqref{eq: effective transition rates free field} and $\mathcal{F}^m_\text{eff}$ from Eq.~\eqref{eq: effective transition rates cavity} of the circular Unruh transition are plotted against $\tilde{\alpha}_m = \alpha/\omega_{mg}$ in Figs.~\ref{fig: Gamma_mag_eff}(a) and \ref{fig: Gamma_mag_eff}(b) for the free-field and the cavity setup, respectively.
	
	Next, we compare these effective transition rates to each other by considering their ratio
	\begin{align}\label{eq: free field vs plates}
		\frac{\Gamma^m_\text{eff}}{\mathcal{F}^m_\text{eff}} = \frac{\omega_{mg}}{c} \frac{L}{2\pi} \frac{\tilde\Gamma^{(3)}_\text{eff}}{\tilde{\mathcal{F}}^{(2)}_\text{eff}}.
	\end{align}
	Based on Eqs.~\eqref{eq: dimensionless effective transition rates free field} and \eqref{eq: dimensionless effective transition rates cavity}, we conclude that this ratio diverges as $\tilde\alpha_m \to \infty$, as illustrated in Fig.~\ref{fig: Gamma_mag_eff}(c).
	At first, the free field setup seems more efficient for observing the circular Unruh transition as $\tilde{\alpha}_m \to \infty$.
	However, since this ratio is also a function of the cavity length $L$, minimizing $L$ allows one to find an interval $[\tilde\alpha_{m,\text{min}}(L), \tilde\alpha_{m,\text{max}}(L)]$ for $\tilde\alpha_{m}$ within which $\Gamma^m_\text{eff} / \mathcal{F}^m_\text{eff} < 1$.
	Crucially, accessing such large values of $\tilde{\alpha}_m$ already poses an experimental challenge in practice, as discussed in Sec.~\ref{Sec: Conclusion} and Table~\ref{tab:physical platforms}.
	In this case, the cavity setup becomes the preferred scheme for measuring the circular Unruh effect at moderate values of $\tilde\alpha_{m}$. 
	Note that the minimal length $L$ should be chosen such that no additional surface effects become relevant, {\it e.\,g.}, the Casimir-Polder force~\cite{Casimir1948, Lamoreaux2005}.
	As an example, in Fig.~\ref{fig: Gamma_mag_eff} we have chosen $L = 5\,\text{\textmu m}$.
	
	\begin{table}[t]
		\caption{\label{tab:physical platforms} Physical platforms for measuring the circular Unruh effect.
			The rows identify three circular setups with which an atom can be confined to a planar circular trajectory, see Fig. \ref{fig:scheme}, where the typical experimental value of the angular velocity $\alpha$ is shown for each setup.
			The columns present the characteristic frequencies of the electric $\omega_{eg}$ and magnetic $\omega_{mg}$ dipole transitions in the atom or ion prepared in the ground or Rydberg state, denoted as $\ket{g}$ in Fig.~\ref{fig: level structure}.
			Various combinations of the circular setup and available internal transitions are evaluated to determine whether the circular Unruh effect can be observed, as marked by "$+$" or "$-$".}
		\centering
		\setlength{\tabcolsep}{4pt}
		
		\resizebox{\columnwidth}{!}{%
			\begin{tabular}{c|c|c|c|c}
				\hline\hline
				& \multicolumn{2}{c|}{Ground state} & \multicolumn{2}{c}{Rydberg state} \\
				Circular setup & $\omega_{eg}$~\cite{Lamb1947} & $\omega_{mg}$~\cite{Reich1956} & $\omega_{eg}$~\cite{Gallagher1994} & $\omega_{mg}$~\cite{Cardman2022} \\
				& ($\sim\mathrm{GHz}$) & ($\sim\mathrm{MHz}$) & ($\sim\mathrm{MHz}$) & ($\sim\mathrm{kHz}$) \\
				\hline\hline
				&  & & & \\
				Ring trap~\cite{Lesanovsky2007, Pandey2019} & - & - & - & - \\
				($\alpha \sim 1\text{--}50$\,Hz) &  & & & \\
				\hline
				&  & & & \\
				Penning trap~\cite{Brown1986, Fan2023} & - & - & - & - \\
				($\alpha \sim$\,kHz) &  & & & \\
				\hline
				&  & & & \\
				Nanoparticle~\cite{Jin2021} & - & + & + & + \\
				($\alpha \sim$\,GHz) &  & & & \\
				\hline\hline
			\end{tabular}%
		}
	\end{table}
	
	\section{Conclusion}
	\label{Sec: Conclusion}
	To summarize, we have investigated the feasibility of detecting the circular Unruh effect for a multilevel atom following a planar circular trajectory coupled to a quantized background electromagnetic field that is initially in its vacuum state in an inertial laboratory frame.
	Our main focus lies on comparing the electric and magnetic dipole transitions in free space and in a cylindrical cavity. 
	In both setups, free space and cavity, the circular Unruh transition rate achieves maximum and minimum values if the dipole matrix element (electric or magnetic), determined by the internal levels in the moving atom, is parallel to or perpendicular to the rotation axis of the atomic circular motion, respectively. 
	
	To achieve a positive effective Unruh transition rate and enable its detection, we have proposed a mitigation scheme that reduces the spontaneous emission rate by a factor of $1/2$ and is based on measuring the ground-state rather than the excited-state population. 
	In the free-field setup, the effective circular Unruh transition rate is shown to be positive when the angular velocity is (much) larger than the corresponding transition frequency $\alpha \gg \omega_{ig}$ for both electric and magnetic dipole transitions.
	However, in the cavity setup, this result only applies to the magnetic dipole transitions.
	This is a direct consequence of the reduction from three (free-field) to effectively two dimensions (cylindrical cavity) and the resulting decrease in the density of modes.
	
	Finally, we examine existing physical platforms for measuring the circular Unruh effect and summarize the results in Table~\ref{tab:physical platforms}.
	The angular velocities so far achieved in ring traps~\cite{Lesanovsky2007} range from $1\,\mathrm{Hz}$ to $50\,\mathrm{Hz}$~\cite{Pandey2019}.
	Consequently, they do not attain the angular velocities $\alpha \gg \omega_{mg}$ required to yield a positive effective circular Unruh transition rate.
	In contrast to ring traps for neutral atoms, Penning traps~\cite{Brown1986} using electrons achieve angular velocities in the $\mathrm{GHz}$ range~\cite{Fan2023}.
	However, the mitigation scheme to suppress spontaneous emission discussed in Sec.~\ref{sec: Introduction} requires a multi-level internal structure of the detector, which necessitates the use of ions.
	This limits the maximum magnetic field strength that can be applied to confine the detector onto a circular trajectory.
	Since the cyclotron frequency scales with the inverse of the mass $m^{-1}$, the angular velocities achievable for hydrogen-like ions in Penning traps ({\it e.\,g.} ${\rm He}^{+}$ at $5\,{\rm G}$) are limited to the $\mathrm{kHz}$ range.
	This regime might satisfy the condition $\alpha \gg \omega_{mg}$ only for magnetic dipole transitions in Rydberg states of the ion.
	The highest angular velocities achieved experimentally so far are in the $\mathrm{GHz}$ regime, realized via nanoparticles~\cite{Jin2021}.
	As such, they obey the condition $\alpha \gg \omega_{mg}$ for both Rydberg atoms and states coupled via magnetic dipole transitions separated by hyperfine splitting.
	Thus, the preferred setup would be a quantum-classical hybrid scheme, in which an optically levitated nanoparticle provides a planar circular trajectory for an atom attached to it, and the electric and magnetic dipole transitions of the atom are measured. 
	
	\begin{acknowledgments}
		We are grateful to W. P. Schleich for his continuing support. 
		We also thank E. Giese, E. P. Glasbrenner, J. Ströhle, and A. Wolf for fruitful and interesting discussions.
		The authors also acknowledge contributions in the form of discussions from the Terrestrial Very-Long-Baseline Atom Interferometry (TVLBAI) proto-collaboration, as well as networking support from COST Actions CA23130 - Bridging high and low energies in search of quantum gravity (BridgeQG) and CA23115 - Relativistic Quantum Information (RQI).
		This work was financially supported by the Science Sphere Quantum Science of Ulm University and by the Center for Integrated Quantum Science and Technology (IQST). 
		The research of the IQST is financially supported by the Ministry of Science, Research and Arts, Baden-W\"urttemberg.
		The QUANTUS project is supported by the German Space Agency at the German Aerospace Center (Deutsche Raumfahrtagentur im Deutschen Zentrum für Luft und Raumfahrt, DLR) with funds provided by the Federal Ministry for Economic Affairs and Climate Action (Bundesministerium f\"ur Wirtschaft und Klimaschutz, BMWK) due to an enactment of the German Bundestag under Grant No. 50WM2450D.
	\end{acknowledgments}
	
	\section*{Author Declarations}
	\subsection*{Conflict of Interest Statement}
	The authors have no conflicts to disclose.
	
	\subsection*{Author Contributions }
	\noindent{\sffamily\small\textbf{Gregor Janson}} Conceptualization (equal); Formal analysis (lead); Validation (equal); Investigation (lead); Methodology (equal);  Visualization (lead); Writing - Original Draft (lead); Writing - Review \& Editing (equal).
	\noindent{\sffamily\small\textbf{Fabio Di Pumpo}} Conceptualization (equal); Formal analysis (support); Validation (equal); Investigation (support); Methodology (equal); Visualization (support); Writing - Original Draft (support); Writing - Review \& Editing (equal); Supervision (equal).
	\noindent{\sffamily\small\textbf{Lorenz Thoma}} Formal analysis (support); Validation (support); Investigation (support); Methodology (support); Writing - Review \& Editing (support).
	\noindent{\sffamily\small\textbf{Maxim Efremov}}  Conceptualization (equal); Formal analysis (support); Validation (equal); Investigation (support); Methodology (equal); Visualization (support); Writing - Original Draft (support); Writing - Review \& Editing (equal); Supervision (equal).

	\section*{Data Availability}
	
	The data that support the findings of this study are available from the corresponding author upon reasonable request.
	
	\appendix
	
	\section{Free-space Unruh transition rate and spontaneous emission rate}\label{app: transition rates}
	In this Appendix we derive the analytical expressions for the rate $\Gamma^i_+$ of the circular Unruh transition and the spontaneous emission rate $\Gamma^i_-$ in the free-field setup via electric ($i = e$) or magnetic ($i = m$) dipole interactions.
	In both cases, the electromagnetic field is initially prepared in the vacuum state $\ket{0}$ defined in the laboratory reference frame.
	The Unruh transition rate $\Gamma^i_+$ is defined as the probability $P^i_+(T)$ per interaction time $T$ for the atom to transition from its internal ground state $\ket{g}$ to the excited state $\ket{i}$, while emitting a photon of arbitrary polarization and frequency in the long-time limit $T \to \infty$, corresponding to Fermi's golden rule.
	The spontaneous emission rate $\Gamma^i_-$ is the inverse process, i.\,e., starting from the excited state $\ket{i}$ the atom decays into its ground state $\ket{g}$.
	
	The probability to find the atom in the corresponding internal state and a photon with wave vector $\vec{k}$ and arbitrary polarization is given by
	\begin{align}\label{eq: App probability 1}
		\begin{split}
			P^{i}_\pm(\vec{k},T)
			=& \frac{1}{(2\pi)^{3}}\sum_{\lambda = 1}^2 \left\vert \int_0^T\dd t \; \Omega^{i}_{\lambda}(\vec{k})\ee^{\ii [\omega_\pm(k)t - \vec{k}\cdot\vec{r}(t)]} \right\vert^2 \\
			=& \frac{1}{(2\pi)^{3}} \vert \mathcal{T}_\pm(\vec{k},T) \vert^2 \sum_{\lambda = 1}^2 \vert \Omega^{i}_\lambda(\vec{k}) \vert^2
		\end{split}
	\end{align}
	with $\omega^{i}_\pm(k)= \omega(k) \pm \omega_{{i}g}$ and
	\begin{align}\label{eq: App T}
		\mathcal{T}_\pm(\vec{k},T)\equiv \int_0^T\dd t\,\ee^{\ii [\omega_\pm(k)t - \vec{k}\cdot\vec{r}(t)]}\;, 
	\end{align}
	where we have used the first-order term of the Dyson series~\cite{Dyson1949}.
	Hence, the probability per time $T$ to find a photon with $\vert\vec{k}\vert = k$ is given by integrating 
	\begin{align}\label{eq: Probability spectrum}
		\begin{split}
			& \frac{P^{i}_\pm(k,T)}{T} = \frac{k^2}{(2\pi)^{3}}\int_0^{\pi}\dd\theta_{\vec{k}} \sin\theta_{\vec{k}} \int_0^{2\pi}\dd\varphi_{\vec{k}} \frac{P^{i}_\pm(\vec{k},T)}{T}\\
			&= \frac{k^2}{(2\pi)^{3}} \int_0^{\pi}\dd\theta_{\vec{k}} \sin\theta_{\vec{k}} \int_0^{2\pi}\dd\varphi_{\vec{k}} \frac{\vert \mathcal{T}_\pm(\vec{k},T) \vert^2}{T} \sum_{\lambda = 1}^2 \vert \Omega^i_\lambda(\vec{k}) \vert^2.
		\end{split}
	\end{align}
	over the polar $\theta_{\vec{k}}$ and the azimuthal $\varphi_{\vec{k}}$ angles of $\vec{k}$.
	
	\subsection{Time integration}
	First, we evaluate the time integral given in Eq.~\eqref{eq: App T}. 
	By using the expression of the wave vector $\vec{k}$ in spherical coordinates, 
	\begin{align}\label{eq: k-vector}
		\vec{k} = k [\sin\theta_{\vec{k}}\cos\varphi_{\vec{k}}, \sin\theta_{\vec{k}}\sin\varphi_{\vec{k}}, \cos\theta_{\vec{k}}]^\text{T},
	\end{align}
	and the center-of-mass position of the atom moving on a circular trajectory with constant radius $r$ and angular velocity $\alpha$, 
	\begin{align}
		\vec{r}(t) = r \left[\cos(\alpha t), \sin(\alpha t), 0\right]^\text{T},
	\end{align}
	we obtain the scalar product
	\begin{align}
		\vec{k}\cdot\vec{r}(t) = r k \sin\theta_{\vec{k}} \cos(\alpha t - \varphi_{\vec{k}}).
	\end{align}
	Inserting this result into Eq.~\eqref{eq: App T} and integrating over $t$ yields
	\begin{align}
		\begin{split}
			\mathcal{T}_\pm(\vec{k},T) 
			=& \int_0^T \dd t\, \ee^{\ii \left[\omega_\pm(k) t - r k \sin\theta_{\vec{k}} \cos(\alpha t - \varphi_{\vec{k}})\right]}\\
			=& T\ee^{\ii \omega_\pm(k) T/2} \sum_{n=-\infty}^\infty (-\ii)^n J_n(rk\sin\theta_{\vec{k}}) \\
			\times &\ee^{\ii n (\varphi_{\vec{k}}-\frac{1}{2}\alpha T)}\sinc\left\{\left[n\alpha - \omega_\pm(k)\right]\frac{T}{2}\right\},
		\end{split}
	\end{align}
	where we have used the Jacobi-Anger expansion~\cite{Abramowitz1964}
	\begin{align}
		\ee^{\ii z \cos\phi} = \sum_{n=-\infty}^\infty \ii^n J_n(z) \ee^{\ii n \phi}
	\end{align}
	in terms of the Bessel functions of the first kind, $J_n(z)$.
	Consequently, this procedure results in
	\begin{align}
		\begin{split}
			&\vert \mathcal{T}_\pm(\vec{k},T) \vert^2 = \sum_{n,m = -\infty}^\infty (-\ii)^{n-m} \ee^{\ii (n-m) (\varphi_{\vec{k}}-\alpha T/2)} \times \\
			&\times J_n(rk\sin\theta_{\vec{k}})J_m(rk\sin\theta_{\vec{k}}) \times\\
			&\times T^2 \sinc\left\{\left[n\alpha - \omega_\pm(k)\right]\frac{T}{2}\right\} \sinc\left\{\left[m\alpha - \omega_\pm(k)\right]\frac{T}{2}\right\}.
		\end{split}
	\end{align}
	In the long-time limit, i.\,e., $T \to \infty$, the representation
	\begin{align*}
		\lim_{T\to\infty} T \sinc(xT) = \pi \delta(x)
	\end{align*}
	of the Dirac delta function shows that $m = n$ are the only non-vanishing terms of the sum.
	Consequently, we obtain
	\begin{align}\label{eq: Time integral long-time limit}
		\begin{split}
			\lim_{T\rightarrow \infty}&\frac{\vert \mathcal{T}_\pm(\vec{k},T) \vert^2}{T}\\ 
			&=\lim_{T\rightarrow \infty}\sum_{n = -\infty}^\infty J_n^2(rk \sin\theta_{\vec{k}}) T \sinc^2\left\{\left[n\alpha - \omega_\pm(k)\right]\frac{T}{2}\right\}\\
			&=\frac{2\pi}{c} \sum_{n = -\infty}^\infty J_n^2(rk \sin\theta_{\vec{k}}) \delta\left(k-\frac{n\alpha \mp \omega_{eg}}{c}\right),
		\end{split}
	\end{align}
	where we have used the representation 
	\begin{align*}
		\lim_{T\to\infty} T \sinc^2(xT) = \pi \delta(x)
	\end{align*}
	for the delta function. 
	
	\subsection{Sum over polarizations and integration over azimuthal angle}\label{app: Rabi frequencies}
	
	Next, we calculate the sum over $\lambda$ and the integral over the azimuthal angle $\varphi_{\vec{k}}$ in Eq.~\eqref{eq: Probability spectrum}.
	The wave vector given by Eq.~\eqref{eq: k-vector} defines the two possible polarization vectors
	\begin{subequations}\label{eq: polarization vectors}
		\begin{align}
			\vec{e}_1(\vec{k}) &= [\cos\theta_{\vec{k}} \cos\varphi_{\vec{k}}, \cos\theta_{\vec{k}} \sin\varphi_{\vec{k}}, -\sin\theta_{\vec{k}}]^\text{T},\\
			\vec{e}_2(\vec{k}) &= [-\sin\varphi_{\vec{k}},\cos\varphi_{\vec{k}},0]^\text{T},
		\end{align}
	\end{subequations}
	such that the vectors $\vec{e}_1(\vec{k})$, $\vec{e}_2(\vec{k})$ and $\vec{k}/\vert\vec{k}\vert$ represent a right-handed orthonormal basis in this ordered sequence.
	
	Inserting Eq.~\eqref{eq: polarization vectors} into the definitions, Eq.~\eqref{eq: Rabi frequencies}, of the Rabi coupling strengths and subsequent summation over both polarizations yields
	\begin{align}\label{eq: sum Rabi squares (appendix)}
		\begin{split}
			&\sum_{\lambda = 1}^2 \left\vert \Omega^i_\lambda(\vec{k}) \right\vert^2 = \left\vert-\Omega^i_x(k) \sin\varphi_{\vec{k}} + \Omega^i_y(k) \cos\varphi_{\vec{k}}\right\vert^2\\
			&+ \left\vert\cos\theta_{\vec{k}}\left[\Omega^i_x(k) \cos\varphi_{\vec{k}} + \Omega^i_y(k) \sin\varphi_{\vec{k}}\right] - \Omega^i_z(k) \sin\theta_{\vec{k}}\right\vert^2.
		\end{split}
	\end{align}
	
	According to Eq.~\eqref{eq: Time integral long-time limit}, the term $\vert \mathcal{T}_\pm(\vec{k},T) \vert^2/T$ is independent of the azimuthal angle $\varphi_{\vec{k}}$.
	This fact allows us to take this term from the integral over $\varphi_{\vec{k}}$ in Eq.~\eqref{eq: Probability spectrum}, to obtain
	\begin{align}\label{eq: Rabi vector sum}
		\begin{split}
			&\int_0^{2\pi}\dd\varphi_{\vec{k}} \sum_{\lambda = 1}^2\left\vert \Omega^i_\lambda(\vec{k}) \right\vert^2\\
			=& \pi \vert \vec{\Omega}^i(k) \vert^2 \left[\sin^2\theta_{\vec{\Omega}^i} \left(1 + \cos^2\theta_{\vec{k}}\right) + 2\cos^2\theta_{\vec{\Omega}^i} \sin^2\theta_{\vec{k}}\right].
		\end{split}
	\end{align}    
	Here, $\theta_{\vec{\Omega}^i}$ is the angle between the Rabi vector $\vec{\Omega}^i(k)$ and the $z$-axis, being the symmetry axis of the circular motion of the atom as depicted in Fig.~\ref{fig:scheme}. This angle is independent of $k$, it in fact corresponds to the angle between the matrix element $\vec{p}_{eg}$, or $\vec{\mu}_{mg}$, and the $z$-axis, respectively.  
	
	Inserting Eqs.~\eqref{eq: Time integral long-time limit} and~\eqref{eq: Rabi vector sum} into Eq.~\eqref{eq: Probability spectrum} results in
	\begin{align}
		\begin{split}
			\lim_{T\rightarrow\infty}&\frac{P^{i}_\pm(k,T)}{T} = \frac{1}{4\pi c} \sum_{n = -\infty}^\infty k^2 \vert\vec{\Omega}^i(k)\vert^2 \delta\left(k-\frac{n\alpha \mp \omega_{ig}}{c}\right)\\
			&\times\int_0^\pi\dd\theta_{\vec{k}} \sin\theta_{\vec{k}} J_n^2(rk \sin\theta_{\vec{k}})\\
			&\times\left[\sin^2\theta_{\vec{\Omega}^i} \left(1 + \cos^2\theta_{\vec{k}}\right) + 2\cos^2\theta_{\vec{\Omega}^i} \sin^2\theta_{\vec{k}}\right].
		\end{split}
	\end{align}
	
	Thus, the total transition rates are given by
	\begin{align}
		\begin{split}
			&\Gamma^{i}_\pm = \lim_{T\rightarrow\infty}\frac{P^{i}_\pm(T)}{T} = \lim_{T\rightarrow\infty}\int_0^\infty\dd k\frac{P^{i}_\pm(k,T)}{T}\\
			&= \frac{1}{4\pi c}\!\!\!\!\sum_{n = \lceil\pm\frac{\omega_{ig}}{\alpha}\rceil}^\infty\!\!\!\!\left[k^{i,(\pm)}_{n} \vert\vec{\Omega}^{i}(k^{i,(\pm)}_{n})\vert\right]^2\!\!\!\int_{-1}^{1}\!\!\!\dd x J_n^2\left(r k^{i,(\pm)}_{n} \sqrt{1-x^2}\right)\\
			&\times\left[\sin^2\theta_{\vec{\Omega}^i} \left(1 + x^2\right) + 2\cos^2\theta_{\vec{\Omega}^i} \left(1 - x^2\right)\right],
		\end{split}
	\end{align}
	where $k^{i,(\pm)}_{n} = (n\alpha\mp\omega_{ig}) / c$.
	
	\section{Minimal radius in cylindrical cavities}\label{app: minimal radius}
	In this Appendix we derive a condition for the minimal radius of a cylindrical cavity such that an atom trapped on a circular trajectory can detect the circular Unruh effect.
	The symmetry of a cavity of length $L$ and radius $R$ with symmetry axis along the $z$-axis requires the electromagnetic vector potential in the form~\cite{Kakazu1996, Stroehle2024}
	\begin{align}\label{eq: magnetic field cylindrical cavity}
		\hat{\vec{A}}^\text{(C)}(\vec{r},t) = \sum_{s,\sigma} \left[\vec{\mathcal{A}}^\text{(C)}_{s\sigma}(\vec{r})\, \hat{a}_{s\sigma} \ee^{-\ii \omega_{s\sigma} t} + \mathrm{h.c.}\right],
	\end{align}
	where the composed index $s = (n,\eta,\ell)$, with $n \in \mathbb{Z}, \eta \in \mathbb{N}$, and $\ell \in \mathbb{N}_0$, labels the cylindrical cavity modes and $\sigma \in \{1,2\}$ numerates the two possible polarizations.
	
	The (discrete) frequencies are given by the dispersion relation
	\begin{align}
		\omega_{s\sigma} = c k_{s\sigma} = c\sqrt{\left(\frac{\chi_{s\sigma}}{R}\right)^2 + \left(\frac{\ell\pi}{L}\right)^2},
	\end{align}
	where $\chi_{s1} \equiv \chi_{n\eta 1}$ is the $\eta$-th root of the $n$-th Bessel function of the first kind $J_n(x)$~\cite{Abramowitz1964}, i.\,e., $J_n(\chi_{n\eta 1})=0$, whereas $\chi_{s2} \equiv \chi_{n\eta 2}$ is the $\eta$-th root of the derivative of the $n$-th Bessel function of the first kind $J^\prime_n(x)$, i.\,e., $J_n^\prime(\chi_{n\eta 2}) = 0$.
	
	The mode functions in Eq.~\eqref{eq: magnetic field cylindrical cavity},
	\begin{subequations}
		\begin{align}
			\vec{\mathcal{A}}^\text{(C)}_{s1}(\vec{r}) &= \sqrt{\frac{\hbar}{2\omega_{s1}\epsilon_0}}  \nabla \times \nabla \times \vec{e}_z \psi_{s1}, \\
			\vec{\mathcal{A}}^\text{(C)}_{s2}(\vec{r}) &= \ii \sqrt{\frac{\hbar\omega_{s2}}{2\epsilon_0}}  \nabla \times \vec{e}_z \psi_{s2},
		\end{align}
	\end{subequations}
	are defined in terms of the scalar mode functions 
	\begin{subequations}
		\begin{align}
			\psi_{s1} &= c_{s1} J_n\left(\chi_{s1}\frac{r}{R}\right) \ee^{\ii n \phi} \cos\left(\ell \pi \frac{z}{L}\right)\\
			\psi_{s2} &= c_{s2} J_n\left(\chi_{s2}\frac{r}{R}\right) \ee^{\ii n \phi} \sin\left(\ell \pi \frac{z}{L}\right)
		\end{align}
	\end{subequations}
	written in cylindrical coordinates $(r,\phi,z)$.
	The normalization coefficients read
	\begin{align}
		& c_{s1} = \sqrt{\frac{2c^2}{\pi L \alpha_{s1} \chi_{s1}^2 \omega_{s1}^2}}
		&&\text{and}
		&&&c_{s2} = \sqrt{\frac{2}{\pi L \alpha_{s2} \chi_{s2}^2 \omega_{s2}^2}}
	\end{align}
	with
	\begin{align}
		&\alpha_{s1} = J_{n+1}^2(\chi_{n\eta 1})
		&& \text{and}
		&&&\alpha_{s2} = J_n^2(\chi_{n\eta 2}) \left[1-\frac{n^2}{\chi_{n\eta 2}^2}\right].
	\end{align}
	
	Now we consider an atom, which is initially prepared in the ground state $\ket{g}$ and moves on a circular trajectory with constant radius $r$ and angular velocity $\alpha$, $(r,\phi,z) = (r_0, \alpha t,z_0)$, while the electromagnetic field is initially prepared in the vacuum state $\ket{0}$ in the laboratory reference frame.
	The probability to find the atom in the excited state $\ket{i}$, with $i=e$ for electric and $i=m$ for magnetic dipole coupling, after the interaction time $T$ is then given by the first order of the Dyson series~\cite{Dyson1949}
	\begin{align}\label{eq: Probability cylindrical cavity}
		\begin{split}
			P_i^\text{(C)}(T) &= \sum_{s\sigma} \left\vert \bra{s\sigma}\otimes\bra{i} \frac{1}{\ii\hbar} \int_{-T/2}^{T/2}\dd t \hat{H}(t) \ket{g}\otimes\ket{0} \right\vert^2,
		\end{split}
	\end{align}
	where the electromagnetic vector potential, Eq.~\eqref{eq: magnetic field cylindrical cavity}, or the corresponding magnetic field $\hat{\vec{B}}(\vec{r},t) = \nabla \times \hat{\vec{A}}(\vec{r},t)$, respectively, are inserted into the Hamiltonian $\hat{H}(t)$, Eq.~\eqref{eq: Hamiltonian magnetic dipole interaction}.
	The relevant term for this transition is exactly given by the counter-rotating term that is typically neglected in the rotating-wave approximation~\cite{Scully1997, CohenTannoudji1998, Schleich2001, Steck2007}.
	
	In the case of constant (dipole) matrix elements $\vec{p}_{eg} = (p_x,p_y,p_z)$ and $\vec{\mu}_{mg} = (\mu_x, \mu_y, \mu_z)$ in the laboratory frame basis, the transition rates consist of summands proportional to
	\begin{align}\label{eq: Delta-function cylindrical cavity z-Direction}
		\frac{T}{2} \sinc\left[\frac{T}{2}(n\alpha - \omega_{ig} - \omega_{s\sigma})\right] \rightarrow \pi\delta(n\alpha - \omega_{ig} - \omega_{s\sigma}),
	\end{align}
	for dipole matrix elements along $z$-direction and
	\begin{align}
		\begin{split}\label{eq: Delta-function cylindrical cavity xy-Plane}
			\frac{T}{2} \sinc&\left\{\frac{T}{2}\left[(n\pm1)\alpha - \omega_{ig} - \omega_{s\sigma}\right]\right\}\\
			&\rightarrow\pi\delta\left[(n\pm1) \alpha - \omega_{ig} - \omega_{s\sigma}\right]
		\end{split}
	\end{align}
	for (dipole) matrix elements in the $xy$-plane, in the long-time limit $T \to \infty$.
	
	Equations~\eqref{eq: Delta-function cylindrical cavity z-Direction} and~\eqref{eq: Delta-function cylindrical cavity xy-Plane} imply a minimal mode number \mbox{$n_\mathrm{min} = \lceil \omega_{ig}/\alpha \rceil \gg 1$} or \mbox{$n_\mathrm{min} = \lceil \omega_{ig}/\alpha \mp 1 \rceil$}, respectively.
	Since $\chi_{n\eta\sigma} > n$ for all mode numbers $(n,\eta,\sigma)$~\cite{Abramowitz1964}, we find the condition $\alpha R > c$ for dipole matrix elements along the $z$-direction~\cite{Zheng2025}, but also the slightly relaxed lower bound
	\begin{align}
		\frac{\alpha R}{c} \geq \frac{\chi_{s\sigma}}{n \pm 1 - \omega_{ig}/\alpha} > \frac{\chi_{s\sigma}}{n+1} \geq \frac{\chi_{112}}{2} \approx 0.92
	\end{align}
	for dipole matrix elements in the $xy$ plane. This is a consequence of the stationary matrix elements rotating with respect to the field modes.
	
	\bibliography{main.bib}
	
	\newpage
	
\end{document}